\documentclass[a4paper,11pt]{article}
\usepackage{jheppub} 

\title{A class of charged black hole solutions in massive (bi)gravity}

\author[a]{Eugeny Babichev}
\author[b,a]{and Alessandro Fabbri}

\affiliation[a]{Laboratoire de Physique Th\'eorique (LPT), Univ. Paris-Sud, CNRS UMR 8627, F-91405 Orsay, France}
\affiliation[b]{Museo Storico della Fisica e Centro Studi e Ricerche Enrico Fermi, Piazza del Viminale 1, 00184 Roma, Italy; Dipartimento di Fisica dell'Universit\`a di Bologna,
Via Irnerio 46, 40126 Bologna, Italy; Departamento de F\'isica Te\'orica and IFIC, Universidad de Valencia-CSIC, C. Dr. Moliner 50, 46100 Burjassot, Spain}

\abstract{
We present a new class of solutions describing charged black holes in massive (bi)gravity.
For a generic choice of the parameters of the massive gravity action, the solution 
is the Reissner-Nordstr\"om-de Sitter metric written in the Eddington-Finkelstein coordinates for both metrics.
We also study a special case of the parameters, for which the space of solutions contains an extra symmetry.
}

\arxivnumber{1405.0581}

\begin{document}

\maketitle
%\flushbottom

\section{Introduction}
During the last few years 
there has been a renewed interest in the theory of massive gravity.
One of the reasons is that there is a version of the massive gravity potential term which evades the generic Boulware-Deser ghost~\cite{deRham:2010ik,Kluson:2011rt}, 
associated with massive gravity theories~\cite{Boulware:1973my} and other higher-order derivative theories~\cite{Ostrogradski}.
Besides, it has been found that the massive gravity theories possess the Vainshtein mechanism~\cite{Vainshtein:1972sx,Babichev:2009us,Koyama:2011xz} (see for a review~\cite{Babichev:2013usa}),
therefore the additional degree(s) of freedom may be hidden effectively to 
pass local gravity tests.
Some issues in massive gravity are debated in \cite{Deser:2012qx,ArkaniHamed:2002sp,Berezhiani:2013dca,Chamseddine:2013lid},
see also the recent review on massive gravity~\cite{deRham:2014zqa}, where these points are addressed.

The first black hole solutions in a generic nonlinear massive gravity were presented long time ago~\cite{Salam:1976as}. 
In the de Rham-Gabadadze-Tolley (dRGT) model of massive gravity (with one fixed Minkowski metric), 
a class of non-bidiagonal Schwarzschild-de-Sitter solutions was presented in~\cite{Koyama:2011xz}.
For a specific choice of the parameters of the massive gravity theory the 
Reissner-Nordstr\"om-de Sitter solution in dRGT model was found in~\cite{Nieuwenhuizen:2011sq}.
For the same choice of parameters another solution describing (uncharged and also charged) spherically symmetric black hole has been found in~\cite{Berezhiani:2011mt}.
In the case of bi-metric massive gravity (when the second metric is also dynamical), the spherically symmetric uncharged black hole solutions were found in~\cite{Comelli:2011wq,Volkov:2012wp}.
See recent reviews on black holes in massive gravity~\cite{VolkovTasinato}.

Black holes in massive gravity have stability properties different from those
in General Relativity (GR), even though all the analytic solutions in massive gravity are related to the GR solutions. 
This happens because
massive gravity possesses extra degrees of freedom. In particular, the bidiagonal Schwarzschild solutions in massive gravity turn out 
to be unstable to radial perturbations~\cite{Babichev:2013una,Brito:2013wya}, even though the rate of instability is extremely slow. 
On the contrary, the non-bidiagonal solutions in massive gravity are stable with respect to this type of
perturbations~\cite{Babichev:2014oua}. 
The stability of the black hole solutions of Ref.~\cite{Berezhiani:2011mt} has been studied in~\cite{Kodama:2013rea} and it was found that the perturbations are identical to those of GR.
This means, in particular, that the dangerous radial mode is absent. 
The absence of the radial mode is connected to a specific choice of the parameters of the Lagrangian~\cite{Babichev:2014oua}.

In this paper we present a new class of charged black hole solutions in (bi)massive gravity, including the dRGT model and also the case with two dynamical metrics.
To find these solutions we write the ansatz of both metrics in the Eddington-Finkelstein form (biEF coordinates)--- a `trick' we used previously in~\cite{Babichev:2014oua} to study perturbations of 
non-bidiagonal uncharged black hole solutions. 
The use of the biEF coordinates makes the analysis extremely simple and allows us to immediately write down the solutions. 
We also discuss a special choice of the parameters of the Lagrangian, for which some solutions were already considered in the literature. 
We show that in this case the space of solutions is richer, and as special cases we recover previously found solutions.

\section{Charged black holes in bigravity}
We consider the bi-gravity extension of the dRGT model of massive gravity~\cite{deRham:2010ik} with two dynamical metrics $g$ and $f$ 
plus standard electromagnetic field coupled to the g-metric,
\begin{eqnarray}\label{action1}
S &=&  M^2_P\int d^4 x \sqrt{-g} \left(\frac{R[g]}{2} + m^2 {\cal U} [g,f] - m^2 \Lambda_g  \right) 	-\frac14 \int d^4 x \sqrt{-g} F_{\mu\nu}F^{\mu\nu}\nonumber\\
 &+& \frac{\kappa M^2_P}{2}\int d^4 x \sqrt{-f} \left(\mathcal{R}[f] -m^2 \Lambda_f \right)
 \end{eqnarray}
where $R[g]$ and $\mathcal{R}[f]$ are the Ricci scalars for the g- and f-metric correspondingly, $\Lambda_g$ and $\Lambda_f$ 
are the (dimensionless) bare cosmological constants, $m$ is the mass parameter (related to the graviton mass), $\kappa$ is a number parametrising 
the difference in the Planck masses for the metrics, ${\cal U}[g,f]$ is the potential mass term and  
$F_{\mu\nu}=\partial_\mu A_\nu - \partial_\nu A_\mu$ is the standard electromagnetic tensor with $A_\mu$ being the electromagnetic potential.
The interaction potential ${\cal U}[g,f]$ can be expressed in terms of the matrix $\mathcal{K}^\mu_\nu = \delta^\mu_\nu - \gamma^\mu_\nu$, 
where $\gamma^\mu_\nu = \sqrt{g^{\mu\alpha}f_{\alpha\nu}} $. 
The potential $\cal{U}$ contains three terms, 
$$
	\mathcal{U} \equiv  \mathcal{U}_2 + \alpha_3 \mathcal{U}_3 + \alpha_4 \mathcal{U}_4,
$$
where $\alpha_3$ and $\alpha_4$ are parameters of the theory, and each of the pieces reads
\begin{equation}
\label{UdRGT}
\begin{aligned}
	\mathcal{U}_2 &= \frac{1}{2!}\left( [\mathcal{K}]^2 - [\mathcal{K}^2] \right), \\
	\mathcal{U}_3 &= \frac{1}{3!}\left( [\mathcal{K}]^3 -3 [\mathcal{K}] [\mathcal{K}^2] +2[\mathcal{K}^3]\right), \\
	\mathcal{U}_4 &= \det (\mathcal{K}),
\end{aligned}
\end{equation}
where  $[\mathcal{K}]\equiv \mathcal{K}^\rho_\rho$ and 
	$[\mathcal{K}^n]\equiv  (\mathcal{K}^n)^\rho_\rho $.

The variation of the action with respect to the metrics $g$ and $f$ gives the modified Einstein-Maxwell equations,
\begin{eqnarray}
	G^{\mu}_{\phantom{\mu}\nu}  &=&  m^2\left( T^{\mu}_{\phantom{\mu}\nu} -  \Lambda_g \delta^\mu_\nu \right) 
		+ \frac{1}{M_P^2} \left( F^{\mu\alpha} F_{\nu\alpha}-\frac14\delta^\mu_\nu F^2 \right),\label{Eg} \\
	\mathcal{G}^{\mu}_{\phantom{\mu}\nu} & = & m^2 \left(
		 \frac{\sqrt{-g}}{\sqrt{-f}}\frac{ \mathcal{T}^{\mu}_{\phantom{\mu}\nu}}{\kappa} - \Lambda_f \delta^\mu_\nu \right) ,\label{Ef}
\end{eqnarray}
where $G^{\mu}_{\phantom{\mu}\nu}$ and $\mathcal{G}^{\mu}_{\phantom{\mu}\nu}$ are the corresponding Einstein tensors for the  metrics $g$ and $f$,
and the mass energy-momentum tensors are given by
\begin{equation}\label{T}
T^{\mu}_{\phantom{\mu}\nu} \equiv \mathcal{U} \delta^{\mu}_{\nu} - 2 g^{\mu\alpha}\frac{\delta \mathcal{U}}{\delta g^{\nu\alpha}},\; 
\mathcal{T}^\mu_{\phantom{\mu}\nu} = -T^\mu_{\phantom{\mu}\nu} + \mathcal{U}\delta^\mu_\nu.
\end{equation}

We shall write down the ansatz for both metrics in the bi-advanced Eddington-Finkelstein form (similar to Ref.~\cite{Babichev:2014oua}, where we
described by a similar ansatz the uncharged non-bidiagonal black holes),
\begin{eqnarray}
ds_g^2 & = & -\left(1-\frac{r_g}{r} + \frac{r_Q^2}{r^2} -\frac{r^2}{l_g^2} \right) dv^2 +2dvdr+r^2 d\Omega^2,\label{metricg}\\
ds_f^2 & = & C^2\left[- \left(1-\frac{r_f}{r}  -\frac{r^2}{l_f^2} \right) dv^2 +2dvdr+r^2 d\Omega^2\right], \label{metricf}
\end{eqnarray}
where $r_g$ and $r_f$ are the Schwarzschild radii for the g- and f-metrics correspondingly, 
$l_g$ and $l_f$ are the ``cosmological'' radii, $C$ is a constant (conformal factor) and $r_Q$ is the length associated with the charge. 
We also take the vector potential in the following form,
\begin{equation}
	A_\mu = \left\{ \frac{Q}{r},0,0,0\right\}
\end{equation}
where $Q$ is the charge of the $g$-black hole. 
With the ansatz (\ref{metricg}), (\ref{metricf}) the Einstein tensors read,
\begin{equation}\label{G}
	G^\mu_{\phantom{\mu}\nu} =
	-\frac{3}{l_g^2}\,\delta^\mu_\nu+
	\frac{r_Q^2}{r^4}\, \text{diag}\left\{  - 1,\, - 1,\,   1,\,  1 \right\},\quad
	\mathcal{G}^\mu_{\phantom{\mu}\nu}  =  -\frac{3}{C^2 l_f^2}\,\delta^\mu_\nu.
\end{equation}
The energy-momentum tensor of the electromagnetic field entering (\ref{Eg}) is given by 
\begin{equation}\label{FF}
F^\mu_{\phantom{\mu}\alpha} F_\nu^{\phantom{\nu}\alpha}-\frac14\delta^\mu_\nu F^2 = 
\frac{Q^2}{2 r^4} \,\text{diag}\left\{ - 1,\,- 1,\,  1,\, 1 \right\}.
\end{equation}
The mass energy-momentum tensor (\ref{T}) for (\ref{metricg}), (\ref{metricf}) gives
\begin{equation}\label{Tmn}
T^{\mu}_{\phantom{\mu}\nu} =
\left(
\begin{array}{cccc}
  \Lambda^{(g)}_m & 0 & 0 & 0 \\
 T^r_{\phantom{r}v} &   \Lambda^{(g)}_m & 0 & 0 \\
 0 & 0 &  \Lambda^{(g)}_m & 0 \\
 0 & 0 & 0 &   \Lambda^{(g)}_m
\end{array}
\right),
\end{equation}
where
\begin{eqnarray}\label{Lg}
	  \Lambda^{(g)}_m =-(C-1)\left(\left(\beta (C-1)^2-3 \alpha  (C-1)+3\right)\right)
\end{eqnarray}
is the effective cosmological constant and 
\begin{eqnarray}\label{Toff}
	T^r_{\phantom{r}v}= - \frac{C}2 \left(\beta  (C-1)^2-2 \alpha  (C-1)+1\right) \left( \frac{r_g-r_f}{r} -\frac{r_Q^2}{r^2} +\frac{r^2}{l_g^2}-\frac{r^2}{l_f^2} \right)
\end{eqnarray}
is the only non-diagonal term.  In the above we introduced the definitions,
\begin{equation}\label{ab}
\alpha\equiv 1+\alpha_3,\ \beta \equiv \alpha_3+\alpha_4.
\end{equation}
The mass energy-momentum for the metric $f$ can be easily found then from (\ref{Tmn}) and (\ref{T}).

Now, from (\ref{G}), (\ref{FF}), (\ref{Tmn}), (\ref{Lg}) and (\ref{Toff}) it is not difficult to see that the Einstein equations~(\ref{Eg}) and (\ref{Ef})
are satisfied when,
\begin{equation}\label{cnd}
\begin{aligned}
\beta  (C-1)^2-2 \alpha  (C-1)+1 &= 0,\\
\sqrt{2}M_P r_Q & = Q,\\
(C-1)\left(\left(\beta (C-1)^2-3 \alpha  (C-1)+3\right)\right) +  \Lambda_g & = \frac{3}{m^2 l_g^2} ,\\
 -\frac{1}{\kappa C^3}\left( C^3 (1-\alpha+\beta) -3C^2 \beta+ 3C (\alpha+\beta)-2\alpha-\beta -1\right) +  \Lambda_f & = \frac{3}{C^2 m^2 l_f^2} . 
\end{aligned}
\end{equation}
In particular, from~(\ref{cnd}) one can find the parameters of the metrics (\ref{metricg}), (\ref{metricf}) in terms of the parameters of the action~(\ref{action1}).

The above results apply for the case of bi-gravity, i.e. when both metrics are dynamical. It is, however, not difficult to formulate the results for the case of one dynamical metric, 
in particular, for the original dRGT theory of massive gravity, when the metric $f$ is flat.
In this case the last two pieces in (\ref{action1}) are absent, i.e. the Einstein-Hilbert term for the $f$-metric and the bare cosmological constant $\Lambda_f$.
Since there is no dynamics for the $f$ metric, we can keep it in the form~(\ref{metricf}), it defines the non-dynamical fiducial background for the theory. 
The metric $g$, on the other hand, is also given by the same ansatz, Eq.~(\ref{metricg}), with the first three conditions of (\ref{cnd}), the last condition of (\ref{cnd}) being absent
in the case of one dynamical metric. 

\section{The case $\beta = \alpha^2$}

There is a special combination of parameters of the action, namely $\beta = \alpha^2$,
for which the space of solutions is much wider than for the general case. 
Indeed, instead of the ansatz (\ref{metricg}) and (\ref{metricf}),  consider now the following very general ansatz,
\begin{eqnarray}
ds_g^2 &=&  -g_{vv} dv^2 +2 g_{vr} dvdr +g_{rr} dr^2 +r^2 d\Omega^2,\label{gs} \\
ds_f^2 & = & C^2\left[-f_{vv} dv^2 +2 f_{vr} dvdr +f_{rr} dr^2 +r^2 d\Omega^2\right], \label{fs}
\end{eqnarray}
where $g_{vv}$, $g_{vr}$, $g_{rr}$, $f_{vv}$, $f_{vr}$, $f_{rr}$ are functions of both $v$ and $r$.
With the ansatz (\ref{gs}) and (\ref{fs}) the matrix $\gamma^\mu_\nu$ has the following form,
\begin{equation}
\label{gammas}
\gamma^{\mu}_{\phantom{\mu}\nu} =
\left(
\begin{array}{cccc}
  \gamma^v_v & \gamma^v_r & 0 & 0 \\
  \gamma^r_v &  \gamma^r_r & 0 & 0 \\
 0 & 0 & C & 0 \\
 0 & 0 & 0 & C
\end{array}
\right),
\end{equation}
where the elements of the matrix $\gamma$ can be straightforwardly expressed in terms of $g_{\mu\nu}$ and $f_{\mu\nu}$.
One can calculate explicitly $T^{\mu}_{\phantom{\mu}\nu}$ in terms of the components of (\ref{gammas}).
Remarkably, applying the conditions
\begin{equation} \label{cond}
\beta=\alpha^2,\, C=1+ \frac{1}{\alpha},
\end{equation}
the mass energy-momentum tensors take very simple forms, 
\begin{equation}\label{Leff}
T^{\mu}_{\phantom{\mu}\nu} = -\frac{1}{\alpha}\, \delta^\mu_\nu,\quad 
\mathcal{T}^{\mu}_{\phantom{\mu}\nu} = \frac{1}{1+\alpha}\frac{\sqrt{-f}}{\sqrt{-g}}  \, \delta^\mu_\nu.
\end{equation}
In the general case, the expressions for the mass energy-momentum tensors are very complicated and we do not write them explicitly here.
Note that the second condition of (\ref{cond}) is the same as the first condition in (\ref{cnd}), 
under the assumption $\beta=\alpha^2$.

We note further that if the Einstein tensors $G^\mu_{\phantom{\mu}\nu}$ and $\mathcal{G}^\mu_{\phantom{\mu}\nu} $ are still given by the same expressions~(\ref{G})
then the Einstein equations (\ref{Eg}) and (\ref{Ef}) are satisfied provided that
\begin{equation}\label{cnd2}
\begin{aligned}
\sqrt{2}M_P r_Q & = Q,\\
\frac{1}{\alpha} +  \Lambda_g & = \frac{3}{m^2 l_g^2} ,\\
 -\frac{1}{1+\alpha} +  \Lambda_f & = \frac{3\alpha^2}{(1+\alpha)^2 m^2 l_f^2} . 
\end{aligned}
\end{equation}
Now, in order for the Einstein tensors to have the desired form~(\ref{G}), the metric $g$ 
should be the Reissner-Nordstr\"om-de Sitter solution, however not necessarily written in the standard form~(\ref{metricg}).
In fact, the ansatz (\ref{gs}) allows us to make in (\ref{metricg})
an arbitrary non-singular coordinate change of the form
\begin{equation}\label{change}
v\to v(v,r),\, r\to r(v,r).
\end{equation}
The same is true for the metric $f$: we can take the metric (\ref{metricf}) and make (in general different) coordinate change~(\ref{change}).

Such a change of coordinates is something trivial in the framework of General Relativity. However, it is not so in the case of massive gravity. 
Indeed, because one of the diffeomorphisms (in the case of two dynamical metrics) is broken, the 
action (\ref{action1}) is  invariant under change of coordinates, when the two metrics transform simultaneously.
For example, for general choice of $\alpha$ and $\beta$, Eqs.~(\ref{metricg}) and (\ref{metricf}) form a solution provided that (\ref{cnd}) are satisfied.
One can make an arbitrary change of coordinates (in particular, as in (\ref{change})), to bring (\ref{metricg}) in a different form, however, at the same time
one must apply the same coordinate change in (\ref{metricf}) also. Only in this case the transformed metrics $g$ and $f$ remain %to be 
a solution. Thus, similar to GR, 
one should treat all the solutions obtained from (\ref{metricg}), (\ref{metricf}) by the simultaneous transformation of both metrics as the same solution, written in a different gauge.

The case $\beta=\alpha^2$, on the other hand, is very different. One still can do the gauge transformation, common for both metrics, as in the general case. 
Apart from this, however, one can also do {\it independent} coordinate change of the form (\ref{change}) for each metric and the resulting expressions for the metrics
will be again a solution. 
Thus this case contains a much richer family of solutions than for the generic choice of $\alpha$ and $\beta$ described in the previous section. 
It seems that the solution for $\beta=\alpha^2$ has an extra symmetry, which is absent in the general case.
The special role of the parameters $\beta=\alpha^2$ has been also noticed in~\cite{Volkov:2012wp}.

The case of one dynamical metric can be easily found from the above consideration. 
Indeed, the solution for the $g$-metric is given by (\ref{gs}), where $g_{vv}$, $g_{vr}$ and $g_{rr}$ are obtained by a coordinate change (\ref{change})
from the metric (\ref{metricg}). The parameters of the metric ansatz should satisfy the first two conditions of (\ref{cnd2}). 
The metric $f$ is given by (\ref{fs}) with arbitrary $f_{vv}$, $f_{vr}$ and $f_{rr}$. 
In the case of black holes in dRGT gravity (with one dynamical metric), the fact that there is an extra freedom in the solution has been discussed in~\cite{Kodama:2013rea}.

Having the general solution at hand, for $\beta=\alpha^2$, it is interesting to relate it to other solutions presented in the literature before. 
In Ref.~\cite{Berezhiani:2011mt} the Reissner-Nordstr\"om-de Sitter solution in dRGT gravity has been found. 
In their solution, when the $f$-metric has the canonical form $df^2= -dt^2 + dr^2 + r^2 d\Omega^2$, 
the $g$-metric is the Reissner-Nordstr\"om-de Sitter solution in the 
Gullstrand-Painlev\'e coordinates, up to the rescaling of the radial coordinate. 
It is not difficult to see that imposing~(\ref{cond}) and choosing $f_{vv}=f_{rr}=1/C^2$, $f_{vr}=0$ and also
the $g$-metric (\ref{gs}) to be the Reissner-Nordstr\"om-de Sitter solution written in Gullstrand-Painlev\'e coordinates, namely,
$$
ds^2 = -dt^2 + \left( dr + dt\,\sqrt{\frac{r_g}{r} - \frac{r_Q^2}{r^2} +\frac{r^2}{l_g^2} } \right)^2 + r^2 d\Omega^2,
$$
one recovers the charged black hole solution found in~\cite{Berezhiani:2011mt}.

Another example of the general solution (\ref{gs}) and (\ref{fs}) for $\alpha^2=\beta$ has been found in \cite{Nieuwenhuizen:2011sq}. 
In \cite{Nieuwenhuizen:2011sq} a bi-diagonal solution has been presented, with $f$ non-dynamical and having the canonical form, 
and the $g$-metric reads
$$
ds^2 = - \left(1-\frac{r_g}{a_1 r} + \frac{r_Q^2}{a_1^2 r^2} -\frac{a_1^2 r^2}{l_g^2} \right) dt^2 + \frac{a_1^2 dr^2}{1-\frac{r_g}{a_1 r} + \frac{r_Q^2}{a_1^2 r^2} -\frac{a_1^2 r^2}{l_g^2}} + a_1^2 r^2 d\Omega^2,
$$
where $a_1$ is connected to the parameters of the action as $a_1 = \frac{\alpha}{1+\alpha}$ in our notations.
It is not difficult to see that by the coordinate change $r\to r/a_1$ one brings this solution to the form (\ref{gs}), (\ref{fs}), with (\ref{cond})
and $g$ being the canonical Reissner-Nordstr\"om-de Sitter form.

\section{Conclusions}
In this paper we presented a new class of charged black hole solutions in massive (bi-)gravity theory.
To find the solution for the general parameters of the action we have written both metrics in the  advanced Eddington-Finkelstein 
coordinates (\ref{metricg}) and (\ref{metricf}). This simple form allows to easily establish that the given ansatz for the metrics is 
indeed a solution once the conditions~(\ref{cnd}) of the parameters of the ansatz are satisfied.
In the case of one dynamical metric (in particular, when $f$ is Minkowski), the solution is also valid, 
the only difference is that in this case the last condition of (\ref{cnd}) does not exist.

We also separately considered the special case of the action parameters  $\beta=\alpha^2$, where $\alpha$ and $\beta$
are given in terms of the parameters of the Lagrangian by (\ref{ab}).
In this case, provided that $C=1+ \frac{1}{\alpha}$,
any choice of the metrics of the form (\ref{gs}) and (\ref{fs})
satisfying Einstein-Maxwell and Einstein equations with a cosmological constant is a solution. 
As particular cases we recovered the solutions found in \cite{Nieuwenhuizen:2011sq} and  \cite{Berezhiani:2011mt}.

{\it Acknowledgments.} The work of E.B. was supported in part by the Grant No. RFBR 13-02-00257-a.

\end{document}